\documentclass[12pt,a4paper,fleqn]{article} 
\usepackage[fleqn]{amsmath} 
\usepackage{amssymb} 
\usepackage{physics} 
\usepackage{graphicx} 
\usepackage{caption} 
\usepackage[bottom]{footmisc} 
\usepackage{cite} 
\usepackage[pdftex]{hyperref} 
\usepackage{ifthen} 
\newboolean{Doppler} 
\newboolean{Twins} 

\begin{document} 

\title{ 
\textbf{Aspects of clock synchronization in \\
relativistic kinematics -- a tutorial} 
} 
\author{ 
\textit{Winfried A. Mitaroff}\,\footnote{ 
\href{mailto:winfried.mitaroff@oeaw.ac.at}
{\,\texttt{winfried.mitaroff@oeaw.ac.at} }  
} \\
Institute of High Energy Physics, \\
Austrian Academy of Sciences, Vienna 
}
\date{16 June 2023} 


\maketitle
\thispagestyle{empty}

\begin{abstract} 

This tutorial, addressing physics teachers and undergraduate students, 
aims at clarifying some aspects of time in special relativity. 
 
In particular, time dilation is usually presented
only as the well-known ratio of lab time over proper time, $R$. 
Above ratio is useful, e.g. for describing the decay length of an unstable
particle in the lab. But essential characteristics of time
dilation, missed out in many textbooks, are manifest in the differences 
(quadratic, absolute and relative) between lab time and proper time. 

After an introduction to the basics, we regard 
the movement of a clock over a fixed spatial distance $\Delta x$. 
Explicit formulae are given for $R$ and for the differences, and their behaviour 
as functions of the velocity is discussed and plotted. 
Approximate formulae are given for very slow movements. 

Time synchronization within an inertial system by "slowly" moving a clock
over a finite distance $\Delta x$, mentioned vaguely in some textbooks,
may quantitatively be comprehended by requiring the time difference to be less
than the clock's time resolution. 
We derive explicit formulae for the maximal velocity obeying these constraints 
(to the best of our knowledge not presented elsewhere). 

\end{abstract} 

\newpage 

\section{Introduction}\label{intro} 

A central feature of relativistic kinematics, in contrast to Newtonian, is the flow of time 
being different when measured in different inertial systems moving relatively to each other. 
This is common textbook knowledge, see e.g. \cite{feynman, berkeley, sexl, inverno}. 

Let $\mathcal{S}$ be an inertial system, called ``laboratory frame'', with spacetime vectors 
$x = ( c \, t, \vec{x} \, )$ obeying a $( + 1, - 1, -1, -1 )$ Minkowski metric, i.e. 
\begin{equation}\label{eq1} 
x^2 = (c \, t)^2 - \vec{x} \, ^2 
\end{equation} 
\indent 
It is trivial to measure the distance $\Delta x = | \vec{x}_2 - \vec{x}_1 |$ between two space 
points $P_1  \, (\vec{x}_1 )$ and $P_2  \, (\vec{x}_2 )$ by means of standard rods at rest, 
and the time interval $\Delta t$ between two events at the \textit{same} space point 
by means of a standard clock.\footnote{ 
The standard rods and clocks are assumed to be calibrated against whatever definition 
is used for the units of length and time, respectively.}

It is less trivial to synchronize two clocks located at \textit{different} space points $P_1$ 
and $P_2$. A correct way to do so relies on the constant speed of light, $c \,$: 
the clock at $P_1$ emits a signal at say $t = 0$, which when received at $P_2$ 
triggers that clock to be set to $t = \Delta x / c$. 
Thus, the clock at $P_2$ gets synchronized with the one at $P_1$. 
 
An alternative way is suggested in some textbooks, e.g. \cite{berkeley, sexl}: 
first synchronize two clocks at $P_1$, then move one of them ``sufficiently slow'' to $P_2$ 
such that the time difference caused by relativistic time dilation may be neglected. 
Upper limits of ``how slow'' this procedure needs to be performed 
are discussed in sect. \ref{slow}. 

Next, let $\mathcal{S'}$ be another inertial system, defined by an object moving with velocity 
$\vec{v} = ( v, 0, 0 )$ w.r.t. $\mathcal{S}$ and called the object's ``rest frame'', with 
spacetime vectors $\xi = ( c \, \tau, \vec{\xi} \, )$ obeying the same Minkowski metric 
as above for $\mathcal{S}$, viz. 
\begin{equation}\label{eq2} 
\xi^2 = (c \, \tau)^2 - \vec{\xi} \, ^2 
\end{equation} 
and with the time $\tau$ called the object's \textit{proper time}. 

The synchronization of two clocks in \textit{different} inertial systems, 
e.g. one in $\mathcal{S}$ and the other in $\mathcal{S'}$, 
is only possible when their positions coincide at a \textit{common space point}. 
If this is the case, then synchronization becomes trivial. 

\section{Time dilation}\label{relat} 

Let w.l.o.g. the object rest in $\mathcal{S'}$ at the spatial origin $O' ( 0, 0, 0  )$, 
and is moving with $v > 0$ along the $x$-axis of $\mathcal{S}$. 
In $\mathcal{S}$, besides the spatial origin $O ( 0, 0, 0  )$, we mark a second 
space point $P ( \Delta x, 0, 0  )$ with $\Delta x > 0$ on its $x$-axis. 

Now regard two events $\mathcal{E}_1$ and $\mathcal{E}_2$, 
occurring in both $\mathcal{S}$ and $\mathcal{S'}$: 
\begin{itemize} 
\item 
$\mathcal{E}_1$ when $O'$ passes through $O$, 
resetting both clocks to $t = 0$ and  $\tau = 0$: \\
the coordinates are $x_1 = ( 0, 0, 0, 0 )$ in $\mathcal{S}$ and 
$\xi_1 = ( 0, 0, 0, 0 )$ in $\mathcal{S'}$, 
\item 
$\mathcal{E}_2$ when $O'$ passes through $P$, 
clock readouts being $\Delta{t}$ and $\Delta{\tau}$, respectively: \\
the coordinates are $x_2 = ( c \, \Delta t, \Delta x, 0, 0 )$ in $\mathcal{S}$ and 
$\xi_2 = ( c \, \Delta \tau, 0, 0, 0 )$ in $\mathcal{S'}$, 
\end{itemize} 
with the 4-distances squared between $\mathcal{E}_1$ and $\mathcal{E}_2$ being 
\footnote{ 
Terms like $\Delta x^2$ are implicitely understood to represent $(\Delta x)^2$. } 
\begin{equation}\label{eq3} 
( x_2 - x_1 )^2 = ( c \, \Delta t )^2 - \Delta x^2 
\qquad \mathrm{and} \qquad 
( \xi_2 - \xi_1 )^2 = ( c \, \Delta \tau )^2 
\end{equation} 
in $\mathcal{S}$ or $\mathcal{S'}$, respectively. 
\textit{Lorentz invariance} imposes both to be equal: 
\begin{equation}\label{eq4} 
\Delta \tau^2 = \Delta t^2 - \left( \frac{\Delta x}{c} \right)^2 
\qquad \mathrm{with} \qquad 
\Delta x = v \cdot \Delta t, 
\qquad \mathrm{hence \ yielding} 
\end{equation} 
\vspace{-6mm} 
\begin{equation}\label{eq5} 
\Delta t = \gamma \cdot \Delta \tau 
\qquad \mathrm{with} \qquad 
\beta = \frac{v}{c} < 1 
\qquad \mathrm{and} \qquad 
\gamma = \frac{1}{\: \sqrt{1 - \beta^2} \: } > 1 
\end{equation} 
Eqs. (\ref{eq4}, \ref{eq5}) exhibit the relativistic \textit{time dilation}. 
The travelled distance, as observed by the object moving with $v$ along $x$, is given by 
\begin{equation}\label{eq6} 
\Delta x' = v \cdot \Delta \tau = \frac{\Delta x}{\gamma} 
\qquad \mathrm{(relativistic} \ \mathit{length \ contraction)} 
\end{equation} 

\subsection{Discussion}\label{discu} 

Intervals of lab time $\Delta t$ are always \textit{greater} than those 
of proper time $\Delta \tau$. 
Time dilation is usually presented only by the ratio of lab time over proper time. 
However, some essential characteristics are manifest in the differences between both. 

\begin{figure}[h!t] 
\centering 
\includegraphics*[width=0.32\textwidth]{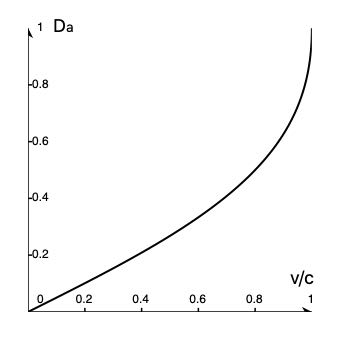} 
\qquad \qquad
\includegraphics*[width=0.32\textwidth]{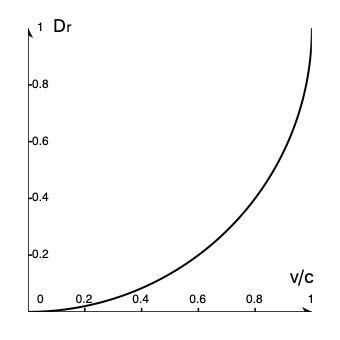} 
\caption[]{ 
Absolute time difference $D_a$ in units [u] over a distance $\Delta x = c \cdot \mathrm{u}$ 
(left), and relative time difference $D_r$ (right), as functions of $\beta = v / c$. } 
\label{fig1} 
\end{figure}

The four quantities defined below expose the effects of time dilation 
for velocities $0 < v < c \,$; these are 
 the ratio $R$, and three ``differences'' $D_q, D_a$ and $D_r$ 
describing the movement over a \textit{fixed distance} $\Delta x$ 
(remember $\Delta t = \Delta x / v$): 
\begin{itemize} 
\item 
The \textit{ratio} $R$ of lab time over proper time, as given by eq. (\ref{eq5}), 
\begin{equation}\label{eq7a} 
R = \frac{\Delta t}{\Delta \tau} = \gamma 
\end{equation} 
which is \textit{increasing} with $v$ from $1$ to $\infty$. 

Thus, an unstable particle of lifetime $\bar{\tau}$ moving at velocity $v$ w.r.t. 
the lab is observed to have a decay length 
$\bar{x} = \beta \, \gamma \cdot c \, \bar{\tau}$ in the lab. 
\vspace{-1mm} \item 
The \textit{quadratic difference} $D_q$ of lab time and proper time, 
see eq. (\ref{eq4}), 
\begin{equation}\label{eq7b} 
D_q = \sqrt{ \Delta t^2 - \Delta \tau^2 \,} = \frac{\Delta x}{c} = \Delta t_L 
\end{equation} 
which is \textit{independent} of $v$, 
is \textit{proportional} to the travelled distance $\Delta x$, and 
is equal to the time $\Delta t_L$ 
which a light signal needs to cross that distance. 

A graphic representation of eqs. (\ref{eq4}, \ref{eq7b}) is shown in the appendix, 
sect. \ref{append}. 
\vspace{-1mm} \item 
The \textit{absolute difference} $D_a$ of lab time and proper time, 
\begin{equation}\label{eq7c} 
D_a = \Delta t - \Delta \tau = 
\frac{\Delta x}{v} \cdot \left( 1 - \frac{1}{\gamma} \right) = 
\frac{\Delta t_L}{\beta} \cdot \left( 1 - \frac{1}{\gamma} \right) 
\end{equation} 
which is \textit{increasing} with $v$ from $0$ to $\Delta x / c$, 
as shown in fig. \ref{fig1} (left). 
\vspace{-1mm} \item 
The \textit{relative difference} $D_r$ of lab time and proper time, 
\begin{equation}\label{eq7d} 
D_r = \frac{D_a}{\Delta t} = \frac{\, \Delta t - \Delta \tau \,}{\Delta t} = 1 - \frac{1}{\gamma} 
\end{equation} 
which is \textit{increasing} with $v$ from $0$ to $1$, 
as shown in fig. \ref{fig1} (right). 
\end{itemize} 

Comparison of eqs. (\ref{eq7b}) and (\ref{eq7c}) exhibits the remarkable fact of the 
quadratic difference~$D_q$ between lab time and proper time being constant, 
hence independent of the velocity~$v$, 
in contrast to the absolute difference $D_a$ between them. 

\subsection{Slow movement}\label{slow} 

For very slow velocities $v \ll c$ (implying $\beta \ll 1$), 
$1 / \gamma = \sqrt{1 - \beta^2 \,}$ of eqs. (\ref{eq7c}, \ref{eq7d}) 
may be approximated to first order in $\beta^2$, resulting in 
\begin{equation}\label{eq7cx} 
D_a \approx \frac{\Delta x}{2 \, c} \cdot \beta = \frac{\Delta t_L}{2} \cdot \beta 
\qquad \mathrm{and} \qquad 
D_r \approx \frac{1}{\, 2 \,} \cdot \beta^2 
\end{equation} 
This linear and parabolic behaviour, respectively, is reflected by the graphs 
of $D_a$ and $D_r$ in fig. \ref{fig1} (left, right) for small values of $\beta$. 

Looking at eqs. (\ref{eq7c}, \ref{eq7d}), eq. (\ref{eq7cx}) or fig. \ref{fig1}, 
the limit $v \rightarrow 0$ is 
\begin{equation}\label{eq7e} 
\lim_{\, v \to 0} D_a = 0 
\qquad \mathrm{and} \qquad 
\lim_{\, v \to 0} D_r = 0 
\end{equation} 
thus, the procedure mentioned in sect. \ref{intro} of synchronizing clocks within 
an inertial system by moving one clock ``slowly'' over a finite distance $\Delta x$ 
is feasible. Since $\lim_{\, v \to 0} \Delta t  = \infty$, 
up to which velocity $v > 0$ may time dilation be neglected, i.e. which is 
the maximal velocity $v \le v_{max}$ such that $D_a$ does not exceed 
the clock's  \textit{absolute time resolution} $\sigma = \sigma (\Delta t)$? 
Applying the constraint $D_a \le \sigma$ to eqs.(\ref{eq7c}, \ref{eq7cx}) yields 
\begin{equation}\label{eq7f} 
v_{max} = \frac{2 \, \sigma \cdot \Delta x}{\, ( \Delta x / c )^2 +\sigma^2 \,} 
\approx \frac{2 \, c^2}{\Delta x} \cdot \sigma 
\end{equation} 

Asking which is $v\,'_{max}$ such that $D_r$ does not exceed the clock's 
\textit{relative time resolution} $\epsilon = \sigma( \Delta t) / \Delta t$: 
apply the constraint $D_r \le \epsilon$ to eqs.(\ref{eq7d}, \ref{eq7cx}), yielding 
\begin{equation}\label{eq7g} 
v\,'_{max} = c \cdot \sqrt{2 \, \epsilon - \epsilon^2 \,} 
\approx c \cdot \sqrt{2 \, \epsilon \,} 
\end{equation} 

The approximations of eqs. (\ref{eq7f}, \ref{eq7g}) are valid only if 
$v_{max}$ resp. $v\,'_{max} \ll c$. Hence, for 
larger time resolutions $\sigma$ or $\epsilon$, the exact formulae must be used. 

\section{Appendix}\label{append} 

Time dilation, see sect. \ref{relat}, may be graphically displayed in a diagram of the 
moving object's proper time $\tau$ vs. the scaled travel distance $x / c$ in the laboratory, 
where eqs.~(\ref{eq4},~\ref{eq7b}) are represented by an orthogonal triangle. 

\begin{figure}[h!t] 
\centering 
\includegraphics*[width=0.55\textwidth]{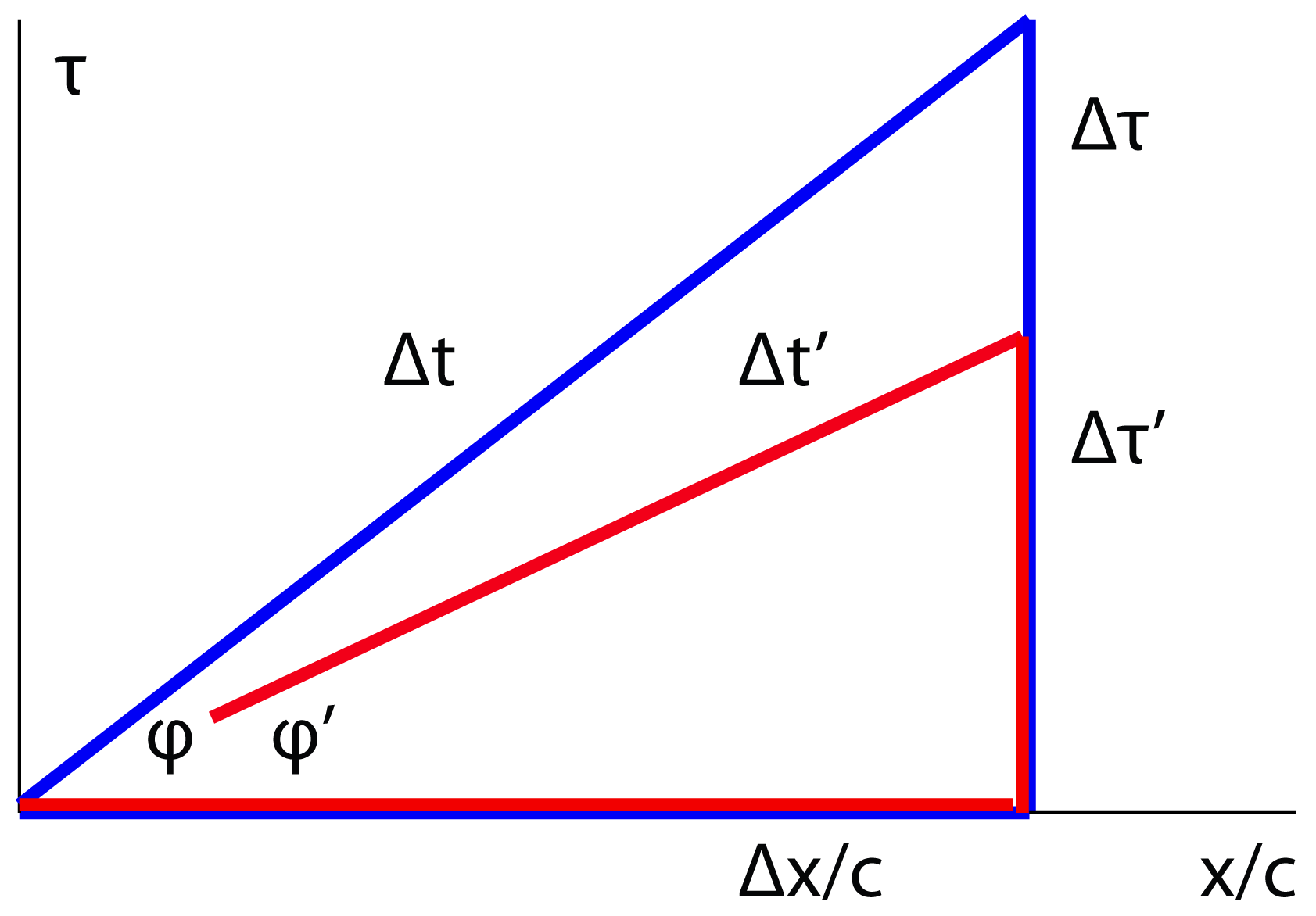} 
\caption[]{ 
Representation of time dilation (red is higher velocity than blue). } 
\label{fig2} 
\end{figure}

This is shown in fig. \ref{fig2} for two velocities $v$ (blue) and $v' > v$ (red), both for 
travelling the \textit{same distance} $\Delta x = v \cdot \Delta t = v' \cdot \Delta t'$. 

Looking first at the blue triangle only, we remark 
\begin{equation}\label{eq14} 
\sin \varphi = \frac{\Delta \tau}{\Delta t} = \frac{1}{\gamma} = \sqrt{1 - \beta^2 \,} < 1 
\qquad \: \: \mathrm{(equivalent \ to \ eqs. \ (\ref{eq5}, \ref{eq7a}))} 
\end{equation} 
\vspace{-5mm} 
\begin{equation}\label{eq15} 
\cot \varphi = \frac{v}{c} \cdot \frac{\Delta t}{\Delta \tau} = \beta \, \gamma = 
\frac{p \, c}{m \, c^2} \qquad \qquad \mathit{(Lorentz \ boost)} 
\end{equation} 
with mass \footnote{ 
The term ``mass'' is understood as the ``rest mass''; the notion ``relativistic mass'' 
is unnecessary, confusing and should consistently be avoided \cite{okun}. } 
$m$ and momentum $p$ of an object moving freely with velocity $v$, 
and obeying \textit{Einstein's equation}: 
energy $E = \gamma \cdot m \, c^2 = \sqrt{(m \, c^2)^2 + (p \, c)^2 \,}$. 

By comparison, the red triangle's $v' > v$ implies its slope angle $\varphi' < \varphi$, 
the lab time $\Delta t' = \Delta x / v' < \Delta t = \Delta x / v$, 
the proper time $\Delta \tau' < \Delta \tau$, 
and their \textit{ratio} $\Delta t' / \Delta \tau' > \Delta t / \Delta \tau$. 
However, their \textit{quadratic difference} $D_q = 
\sqrt{\Delta t'^2 - \Delta \tau'^2 \,} = \sqrt{\Delta t^2 - \Delta \tau^2 \,} = \Delta x / c$ 
remains the same, in agreement with eq. (\ref{eq7b}) of sect. \ref{discu}. 

The \textit{absolute difference} of lab time and proper time can be expressed as 
\begin{equation}\label{eq16} 
D_a = \Delta t - \Delta \tau = \frac{\, \Delta t^2 - \Delta \tau^2}{\Delta t + \Delta \tau} = 
\frac{(\, \Delta x / c )^2}{\, \Delta t + \Delta \tau \,} 
\end{equation} 
which is equivalent to eq. (\ref{eq7c}). 
The limit $v \rightarrow 0$ (implying $\gamma \rightarrow 1$) corresponds to dragging 
the triangle's upper corner upward to $\infty$, while the base leg $ \Delta x / c$ 
(i.e. the quadratic difference) remains constant; the other leg and the hypothenuse 
grow to $\lim_{\, v \to 0} \Delta \tau = \lim_{\, v \to 0} \Delta t  = \Delta x / 0 = \infty$. 
Thus, eq. (\ref{eq16}) yields 
\begin{equation}\label{eq17} 
\lim_{\, v \to 0} D_a = \frac{(\Delta x / c )^2}{\infty} = 0 
\end{equation} 
which is in agreement with eq. (\ref{eq7e}) of sect. \ref{slow}. 



\section*{Acknowledgements}\label{ackn} 

The author wishes to thank \textit{Wolfgang Lucha} (HEPHY Vienna) and 
\textit{Manfried Faber} (Atominstitut of TU Vienna) 
for helpful suggestions, critical comments, and a careful reading of the manuscript. 

\vspace{10mm}


\end{document}